\documentclass[sigconf,edbt]{acmart-edbt2024}

\def\BibTeX{{\rm B\kern-.05em{\sc i\kern-.025em b}\kern-.08em
    T\kern-.1667em\lower.7ex\hbox{E}\kern-.125emX}}

\usepackage{booktabs} 
\usepackage{subcaption}
\usepackage{enumitem}
\usepackage{multirow}
\usepackage[normalem]{ulem}

\newtheorem{mydef}{Definition}

\setcopyright{rightsretained}

\acmDOI{}

\acmISBN{XXXXXXXXXXXX}

\acmConference[EDBT 2026]{29th International Conference on Extending Database Technology (EDBT)}{24th March - 27th March, 2026}{Tampere, Finland}
\acmYear{2026}

\begin{document}
\title{DistillER: Knowledge Distillation in Entity Resolution with Large Language Models}

\author{Alexandros Zeakis}
\affiliation{%
  \institution{Department of Informatics and Telecommunications, National Kapodistrian University of Athens}
  \city{Athens} 
  \state{Greece} 
}
\email{alzeakis@di.uoa.gr}
\affiliation{%
  \institution{Athena Research Center}
  \city{Athens} 
  \state{Greece} 
}
\email{azeakis@athenarc.gr}

\author{George Papadakis}
\affiliation{%
  \institution{Department of Informatics and Telecommunications, National Kapodistrian University of Athens}
  \city{Athens} 
  \state{Greece} 
}
\email{gpapadis@di.uoa.gr}

\author{Dimitrios Skoutas}
\affiliation{%
  \institution{Athena Research Center}
  \city{Athens} 
  \state{Greece} 
}
\email{dskoutas@athenarc.gr}

\author{Manolis Koubarakis}
\affiliation{%
  \institution{Department of Informatics and Telecommunications, National Kapodistrian University of Athens}
  \city{Athens} 
  \state{Greece} 
}
\email{koubarak@di.uoa.gr}

\renewcommand{\shortauthors}{}

\begin{abstract}

Recent advances in Entity Resolution (ER) have leveraged Large Language Models (LLMs), achieving strong performance but at the cost of substantial computational resources or high financial overhead. 
{\color{black}Existing LLM-based ER approaches operate either in unsupervised settings and rely on very large and costly models, or in supervised settings and require ground-truth annotations, leaving a critical gap between time efficiency and effectiveness. To make LLM-powered ER more practical, we investigate Knowledge Distillation (KD) as a means to transfer knowledge from large, effective models (Teachers) to smaller, more efficient models (Students) without requiring gold labels. We introduce DistillER, the first framework that systematically bridges this gap} across three dimensions: (i) Data Selection, where we study strategies for identifying informative subsets of data; (ii) Knowledge Elicitation, where we compare single- and multi-teacher settings across LLMs and smaller language models (SLMs); and (iii) Distillation Algorithms, where we evaluate supervised fine-tuning and reinforcement learning approaches. Our experiments reveal that supervised fine-tuning of Students on noisy labels generated by LLM Teachers consistently outperforms alternative KD strategies, while also enabling high-quality explanation generation. Finally, we benchmark DistillER against established supervised and unsupervised ER methods based on LLMs and SLMs, demonstrating significant improvements in both effectiveness and efficiency.
\end{abstract}

%
%



\maketitle

\section{Introduction}

Entity Resolution (ER) is the task of detecting different records that refer to the same real-world entity \cite{DBLP:journals/csur/ChristophidesEP21}. It is a fundamental problem in data integration with applications ranging from question answering to analytics \cite{DBLP:journals/pvldb/DongS13,DBLP:series/synthesis/2015Christophides}. 

In practice, ER solutions typically involve two steps \cite{DBLP:series/synthesis/2015Christophides}:
(i) Blocking, which reduces the quadratic comparison space by generating candidate record pairs that are likely to match, and (ii) Matching, a more fine-grained decision task of determining which candidate pairs truly correspond to the same entity. 
The former step has been extensively studied \cite{DBLP:journals/pvldb/ZeakisPSK23, DBLP:journals/pvldb/PaulsenGD23, zeakis2025depth}, with state-of-the-art methods already achieving high effectiveness. Regarding Matching, prior work has approached it with unsupervised methods based on pre-trained small language models (SLMs) \cite{wu2020zeroer,ge2021collaborem}, which avoid labeled data, as well as with supervised methods \cite{Mudgal2018sigmod, DBLP:conf/edbt/BrunnerS20, DBLP:journals/pvldb/0001LSDT20, DBLP:journals/pacmmod/TuFTWL0JG23}, which 
train binary classifiers on labeled examples.

With the rise of Large Language Models (LLMs), recent work adapted them to 
Matching using different prompting strategies. Two dominant settings have emerged \cite{DBLP:journals/corr/abs-2405-16884}, as shown 
in Fig.~\ref{fig:prompt_variations}.
\begin{enumerate}[leftmargin=*]
    \item In the MATCH setting, the model takes a candidate pair as input and answers the binary question: ``Is this pair a match?''. 
    \item The SELECT setting presents the model with a query record and multiple candidates, asking it to ``Select the correct match''.
\end{enumerate}
MATCH has been studied in unsupervised scenarios \cite{peeters2023using, DBLP:journals/corr/abs-2310-11244} as well as supervised ones \cite{DBLP:journals/corr/abs-2409-08185}, where fine-tuned GPT-3.5 has been shown to outperform pre-trained GPT-4. SELECT
yields strong performance in unsupervised mode only with large models such as GPT-3.5 \cite{DBLP:journals/corr/abs-2405-16884}, while ensembles of medium-sized models can achieve competitive accuracy but at a higher computational and time cost \cite{zeakis2025avenger}. 
In supervised settings, fine-tuned medium-sized LLMs (around 8B parameters) have even surpassed pre-trained GPT-3.5, highlighting the efficiency gains of fine-tuning within SELECT \cite{zeakis2025avenger}. 
{\color{black} Despite these advances, current LLM-based ER methods operate in two settings:
(i) unsupervised approaches that rely on very large or proprietary models at high computational and financial cost or (ii) supervised approaches that require manually labeled data. There is currently no systematic solution that transfers the effectiveness of LLMs to efficient models without access to ground-truth labels.}

However, fine-tuning requires labeled data that may not always be available, while large models are computationally expensive and slow to run. A common solution to this, that has been overlooked in ER literature, is Knowledge Distillation (KD) \cite{buciluǎ2006model}, 
a model compression technique, where a smaller \textit{student} model learns to mimic the behavior of a larger \textit{teacher}. Early applications focused on reducing the size of language models, as exemplified by DistilBERT \cite{sanh2019distilbert}. With the advent of LLMs, KD has evolved beyond simple architectural compression to address the broader challenge of transferring knowledge from teachers to students, capturing not only predictions but also reasoning patterns \cite{xu2024survey}. A successful example is the DeepSeek-r1 model~\cite{DBLP:journals/corr/abs-2501-12948}.
However, applying KD to Matching is not straightforward,
raising critical questions: what data should be distilled, how teacher knowledge should be elicited, which algorithms are most effective, etc.


In this work, we introduce DistillER, the first framework for applying KD to ER
and, in particular, to Matching with LLMs, without requiring ground-truth labels.
While KD has been widely explored in NLP, its adaptation to Matching remains underexplored. DistillER addresses this gap by introducing a framework utilizing ER-specific adaptations of distillation methods. It comprises three main components: (i) 
Data Selection, which determines the training data to be used during distillation. 
(ii) Knowledge Elicitation, which depends on the choice of Teacher and the Instructions given—such as whether the model should provide a single answer or an accompanying explanation. The main process here involves the teacher producing labeled data to guide the student. 
(iii)
Distillation Algorithm, which involves choosing the Student model and defining the training strategy. This can take the form of Supervised Fine-Tuning (SFT) or Reinforcement Learning (RL) approaches. We put this framework into practice by evaluating the main instantiations of DistillER against state-of-the-art ER methods leveraging both LLMs and SLMs, across supervised and unsupervised settings. Our thorough experimental study demonstrates consistent gains, with DistillER achieving the best overall average performance.

In short, our contributions are as follows:
\begin{itemize}[leftmargin=*]
\item We introduce DistillER, the first framework for Knowledge Distillation in Entity Resolution with LLMs.
\item We design and evaluate data selection strategies to support effective knowledge transfer.
\item We define a knowledge elicitation process that depends on the choice of teacher and the instructions provided, such as whether the model produces a single answer or an accompanying explanation.
\item We examine the main distillation algorithms, focusing on the selection of the student model and the training strategy, including SFT and RL methods.
\item We perform a comprehensive experimental study, including component-wise analysis and benchmarks against state-of-the-art ER methods on real-world datasets.
\end{itemize}


The rest of this paper is structured as follows: Section~\ref{sec:related} reviews related work and Section~\ref{sec:problem} defines the key concepts of our problem. Section~\ref{sec:framework} outlines the framework of DistillER and its key components. Section~\ref{sec:evaluation} presents our evaluation, while Section~\ref{sec:conclusions} concludes the paper.

\section{Related Work}
\label{sec:related}

We organize related work into three main areas: Entity Matching, which includes approaches based on both SLMs and LLMs, Knowledge Distillation, covering general methods for LLMs as well as their emerging applications in ER and Reinforcement Learning, describing practical optimization strategies.

\subsection{Entity Matching}
\noindent \textbf{Matching with SLMs.} \textit{In the unsupervised setting},
ZeroER \cite{wu2020zeroer} represents candidate pairs with similarity scores across multiple functions and models the resulting vectors with a Gaussian Mixture Model, using adaptive regularization and transitivity to improve accuracy. CollaborEM \cite{ge2021collaborem} adopts a self-supervised strategy that avoids manual annotations by combining automatic label generation with collaborative EM training, capturing both graph- and sentence-level signals. 
\textit{In the supervised setting}, Sudowoodo \cite{wang2023sudowoodo} employs contrastive representation learning to produce similarity-aware representations, enabling effective fine-tuning with minimal labels. HierGAT \cite{yao2022entity} introduces a Hierarchical Graph Attention Transformer that jointly models attribute relationships and matching decisions, enriching contextual embeddings. Finally, Unicorn \cite{DBLP:journals/pacmmod/TuFTWL0JG23} proposes a unified architecture for binary data matching tasks with a generic encoder and mixture-of-experts classifier, supporting cross-dataset knowledge sharing and even zero-shot predictions. We experimentally compare these methods against DistillER in Section \ref{sec:evaluation}.



\vspace{2pt}
\noindent \textbf{Matching with LLMs.}
A key aspect of LLM-based matching 
is prompt engineering, which ranges from zero-shot to few-shot prompting. An extensive study of MATCH prompts is given in \cite{DBLP:journals/corr/abs-2310-11244}, primarily with pre-trained models, and is extended in \cite{DBLP:journals/corr/abs-2409-08185} to include fine-tuning, focusing on cross-domain generalization and training dataset size.

In SELECT prompts, where multiple candidates are evaluated in one query, 
strong performance is achieved with GPT-3.5 in \cite{DBLP:journals/corr/abs-2405-16884}
and with a committee-based approach combining multiple 8B models and voting in \cite{zeakis2025avenger}. In the latter case, 
fine-tuning the 8B models
results in higher performance than pre-trained GPT-3.5, mitigating position bias.

Other works include \cite{DBLP:journals/corr/abs-2405-04820}, which studies augmentation strategies to expand training data, and \cite{DBLP:journals/corr/abs-2409-04073}, which shows that transfer learning with fine-tuning can outperform zero-shot baselines.

We experimentally compare against that state-of-the-art LLM-based Matching algorithms with DistillER in combination with small, open-source models in Section \ref{sec:evaluation}.


\begin{figure}[!t]
\centering
\includegraphics[trim=0 0 0 0, width=1.0\linewidth, height=70mm]{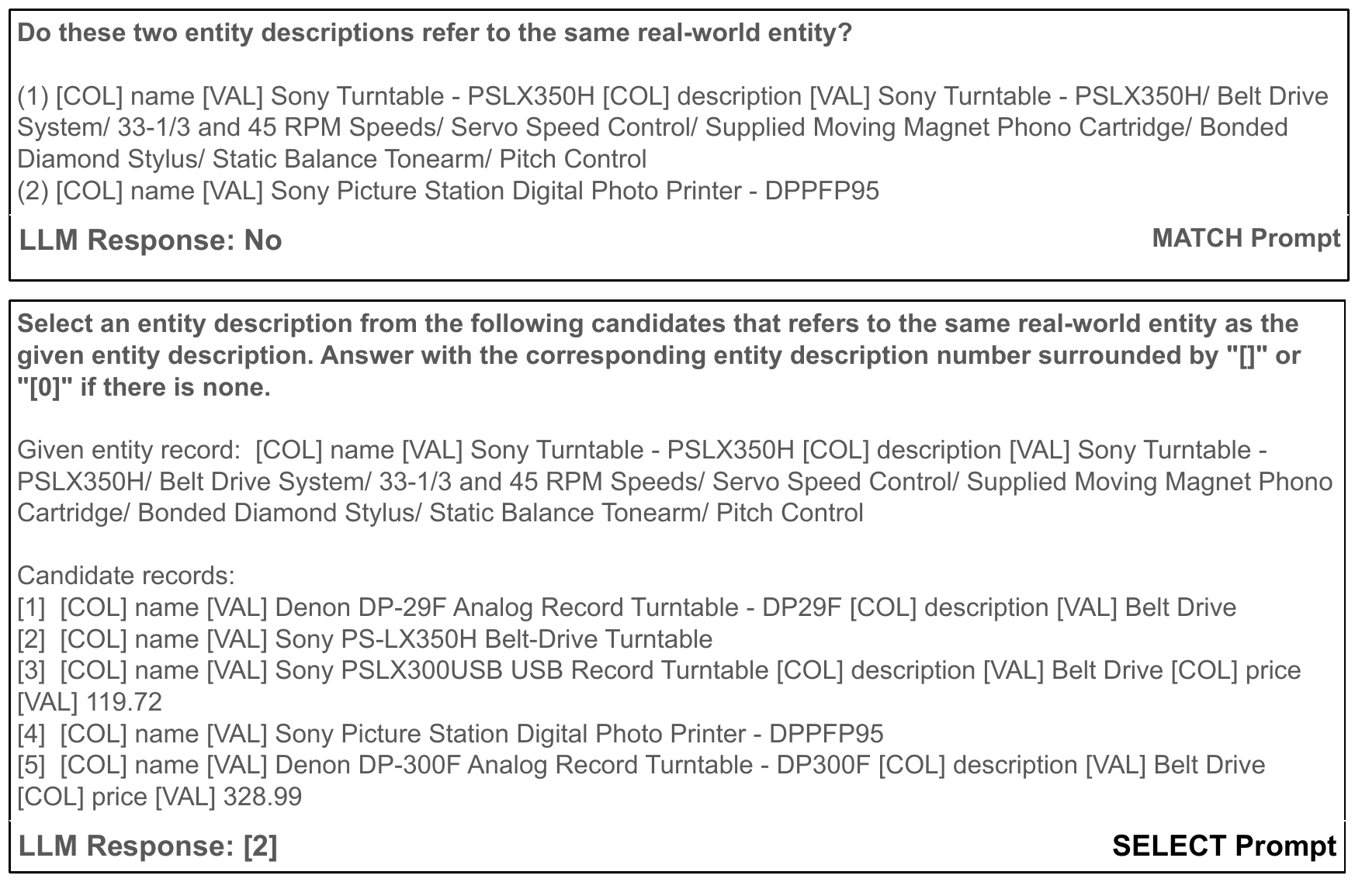}
\vspace{-25pt}
\caption{Example of MATCH vs SELECT prompt.}
\vspace{-20pt}
\label{fig:prompt_variations}
\end{figure}

\subsection{Knowledge Distillation with LLMs}

Existing work on KD with LLMs can be grouped into four main dimensions \cite{xu2024survey}: (i) Knowledge Elicitation, where training data is generated or augmented by querying teacher models; (ii) Distillation Algorithms, ranging from standard supervised fine-tuning to divergence-based, reinforcement learning, or ranking-based approaches; (iii) Student Enhancement, which targets improvements on downstream tasks such as natural language understanding, natural language generation, and information retrieval; and (iv) Vertical Distillation, which adapts KD to domain-specific applications in areas like law, healthcare, and finance. 
\textit{Recently, the emphasis shifted from general compression to task-specific skill transfer, highlighting methods for reasoning, alignment, and domain adaptation.}

In the field of ER, the only relevant work is \cite{DBLP:conf/emnlp/WadhwaKWWK24}, which generates explanations for entity pairs with GPT-3.5 and uses them, along with labels, to fine-tune a smaller model (Flan-T5). Our framework differs by extending instruction-based distillation beyond pairwise setups. In Entity Alignment (EA), \cite{chen2024entity} proposes LLM4EA, a framework combining LLM annotators, active learning, and unsupervised label refinement to produce high-quality labels for fine-tuning at scale. In Entity Linking, \cite{liu2023towards} introduces a Multi-View Enhanced Distillation framework that splits entities into multiple views and distills fine-grained knowledge from cross-encoders to dual-encoders while preserving global structure. Finally, \cite{meduri2020comprehensive} presents a benchmark for active learning in EM, showing that carefully combining classifiers with selection strategies can rival full supervision with fewer labels and greater efficiency.
\textit{Overall, while some distillation efforts have addressed EA and EL, there has been little systematic exploration of ER, especially in studying SFT or RL as core paradigms.}






\subsection{Reinforcement Learning} 
Reinforcement Learning has become a central paradigm for aligning LLMs with preference data. A prominent approach is Reinforcement Learning from Human Feedback (RLHF), where human annotators provide preference labels to train a reward model that guides optimization \cite{schulman2017proximal}. In contrast, Reinforcement Learning from AI Feedback (RLAIF) eliminates the reliance on human annotators by generating preferences using another AI model, which then supervises the student model. Proximal Policy Optimization (PPO) \cite{schulman2017proximal} is a standard reinforcement learning algorithm for both RLHF and RLAIF, using clipped surrogate objectives with regularization to ensure stable policy updates. Recent alternatives include Generalized Reward Policy Optimization (GRPO) \cite{shao2024deepseekmath}, which replaces the value function with group-based advantage estimation for more efficient updates in reasoning-heavy tasks, and Direct Preference Optimization (DPO) \cite{rafailov2023direct}, which bypasses reward modeling altogether by directly aligning the model with pairwise preferences.
We adopt these reinforcement learning strategies within the DistillER framework, integrating GRPO and DPO as alternative optimization methods for the distillation.
\section{Problem Definition}
\label{sec:problem}

We now define the core notions of the task we tackle in this work.

\begin{mydef}[Entity]
We define an entity $e$ as a set of attributes (i.e., name-value pairs) $a_i$, and denote it as 
$e = \{a_1,a_2,.., a_n\}$.
\end{mydef}

\begin{mydef}[Tuple]
Given two sets of entities, $\mathcal{D}_1$ and $\mathcal{D}_2$, consider a query entity $q \in \mathcal{D}_1$ and a set of candidate entities $C \subseteq \mathcal{D}_2$, typically obtained through a Blocking step from another data source.  
We define the \emph{tuple} $t_{q,C}$ as the pair consisting of the query entity $q$ and its associated candidate set $c$, i.e.,
$t_{q,C} = (q,C)$.
\end{mydef}

{\color{black}Tuples are naturally generated when blocking relies on approximate nearest neighbor search with pre-trained language models, as in \cite{DBLP:journals/pvldb/ZeakisPSK23, zeakis2025depth}. Hence, we can estimate the blocking similarity $sim(c)$ between the query entity $q$ and each candidate match $c$ through the cosine similarity of their embedding vectors.}

In this context, we can formalize \emph{Entity Matching (EM)}, i.e.,  
the task of determining whether two entities, possibly coming from different sources, refer to the same real-world object, as follows:

\begin{mydef}[Entity Matching with Knowledge]
Let $\mathcal{T}$ denote a collection of tuples $t_{q,C}$, and let $K$ represent knowledge generated about these tuples (e.g., labels, features, or model predictions).  
The problem of \emph{Entity Matching with Knowledge} is to train a model $M$ on $(\mathcal{T}, K)$ 
to predict whether entities across different data sources
correspond to the same real-world object.
\end{mydef}

\section{Framework}
\label{sec:framework}

\begin{figure*}[!t]
\centering
\includegraphics[trim=0 0 0 0, width=1.0\linewidth]{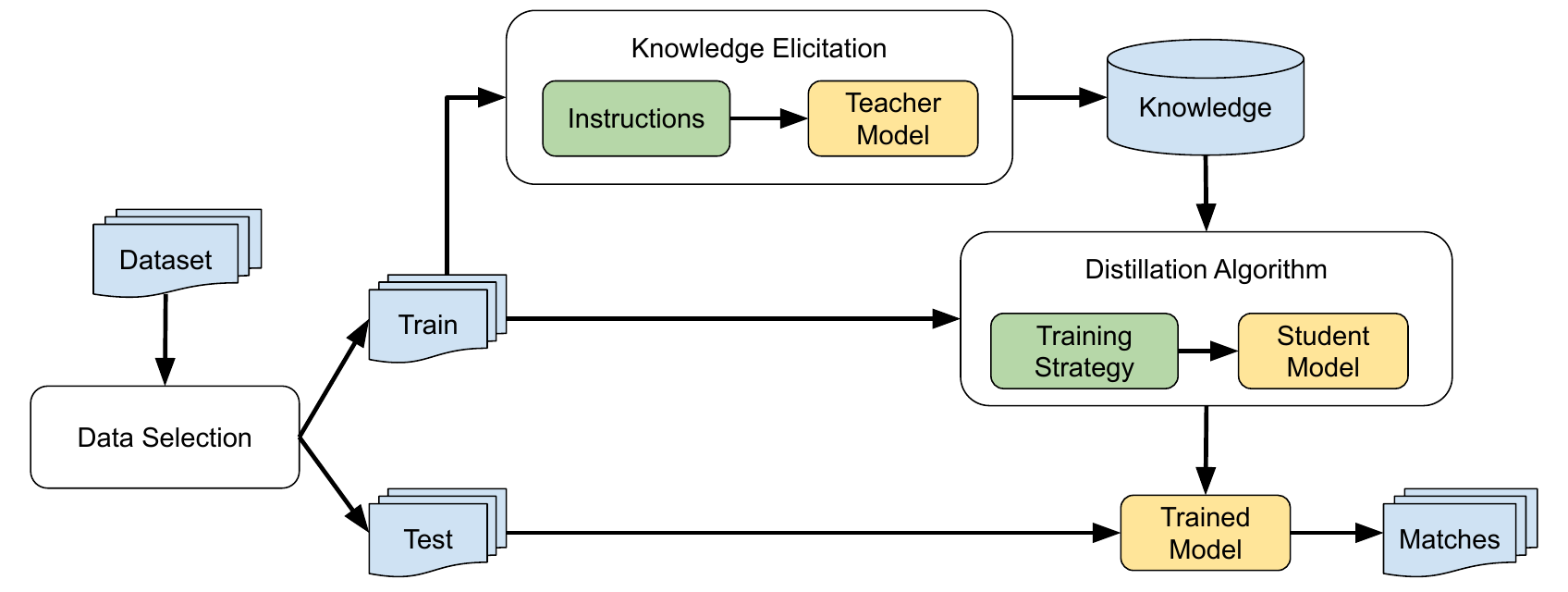}
\vspace{-15pt}
\caption{The framework of DistillER, highlighting the three key components.}
\vspace{-10pt}
\label{fig:framework}
\end{figure*}

We now 
introduce our DistillER framework, which addresses 
Entity Matching with Knowledge through an end-to-end way that leverages self-supervised learning. At its core, a teacher model 
generates the knowledge required for self-training a student model, while the framework also incorporates an effective strategy for splitting the data into training and test sets. As illustrated in Figure \ref{fig:framework}, DistillER consists of three main phases: (i) \textit{Data Selection}, which identifies representative examples to be used in knowledge distillation; (ii) \textit{Knowledge Elicitation}, which extracts and organizes knowledge from teacher models, taking into account different model types and instructions; and (iii) \textit{Distillation Algorithm}, which transfers the elicited knowledge to the student model, focusing on the choice of student architectures as well as the training strategies employed to integrate the distilled knowledge effectively. We elaborate on each of these phases in the following.

\subsection{Data Selection}

This step receives as input a set of tuples from blocking and splits it into two disjoint subsets: the training and the testing one. 
The former is used to generate knowledge from the Teacher model, while the latter is used for evaluating the performance of the Student model. The training set 
consists of two types of tuples: (i) the \textit{positive tuples}, where the actual match of the query entity is included in the candidates, and (ii) the \textit{negative tuples}, where the candidates lack the actual match to the query entity.

{\color{black} In Data Selection, we address two main challenges: (i) how to represent each tuple, and (ii) how to separate positive and negative tuples. For the former, we use the similarity scores generated by blocking, which can either be aggregated into a single value per tuple or represented as a vector by dividing the score range into bins. 
For the latter, 
since we assume no ground-truth labels, we employ two strategies based on the previous tuple representations: (i) The ranking strategy orders tuples by their aggregated score, with the top tuples treated as positives and the bottom ones as negatives. (ii) The clustering strategy groups tuples based on their vector representation into two clusters -- one corresponding to positives and the other to negatives. After either strategy, we randomly select $p$ positive tuples and $n$ negative tuples to form the training set, while the remaining tuples constitute the testing set. 
}

\subsubsection{Ranking}
This approach sorts all tuples based on a computed score and then designates the top $p$ and the bottom $n$ ranked ones as positive and negative, respectively. We define two possible scoring functions for a tuple $t$:

\begin{itemize}[leftmargin=*]
    \item \textbf{Max score:} 
        $r^{\text{max}}_t = \max_{c \in t} \text{sim}(c)$,
    where $\text{sim}(c)$ denotes the similarity score of candidate $c$, as derived from blocking.

    \item \textbf{Top-2 score:} 
        $r^{\text{top-2}}_t = \frac{\text{sim}(c_1) + \text{sim}(c_2)}{2}$,
    where $c_1$ and $c_2$ are the two candidates in $t$ with the highest similarity scores.
\end{itemize}

{\color{black}The max score in a tuple provides a strong indication of whether the correct match is among the candidates. Tuples involving only low similarity scores are unlikely to be positive, in contrast to those with at least one high score.
To reduce the impact of outliers, we also consider the top-2 score, which averages the two highest similarity scores.}

\vspace{2pt}
\noindent
\textbf{Example.} \textit{To illustrate the differences between these two functions, assume that the tuple $t_1$ in the SELECT prompt of Figure \ref{fig:prompt_variations} gets the following similarity scores from blocking for the candidate entities $e_1$-$e_5$: \{0.74, 0.95, 0.90, 0.70, 0.76\}. Assume also another tuple $t_2$ 
with five candidate entities that have the following scores: \{0.42, 0.37, 0.24, 0.39, 0.99\}. The max-scores are $r^{\text{max}}_{t_1}=0.95$ and $r^{\text{max}}_{t_2} = 0.99$, while the top-2 ones are $r^{\text{top-2}}_{t_1}=0.925$ and $r^{\text{top-2}}_{t_2}$ = 0.705.}


\subsubsection{Clustering}

The second strategy treats data selection as a clustering problem with two clusters. {\color{black} 
Unlike similar Active Learning approaches (e.g., \cite{meduri2020comprehensive}), which formulate this task as a classification problem, we address it in an unsupervised way, due to the lack of labelled instances.
We transform the similarity scores of each tuple 
into a histogram vector with a fixed number of $b$ bins. Each bin counts the scores falling within its range, thus providing a balanced representation for all tuples.} The resulting vectors can then be clustered using standard algorithms with the number of clusters set to two. We then label the cluster containing the tuples with the highest similarity scores as \textit{positive}, while the other is treated as \textit{negative}. Finally, we randomly select $p$ and $n$ tuples, respectively.

\vspace{2pt}
\noindent
{\color{black}
\textbf{Example.} \textit{Using the similarity scores of $t_1$ and $t_2$ from the previous example, we map each tuple to a 10-dimensional histogram vector with bins of width 0.1 over $[0,1]$.
For $t_1$, three scores fall into the $[0.7,0.8)$ interval and two into $[0.9,1.0]$, resulting in the vector $[0,0,0,0,0,0,0,3,0,2]$.
For $t_2$, one score falls into $[0.2,0.3)$, two into $[0.3,0.4)$, one into $[0.4,0.5)$, and one into $[0.9,1.0]$, yielding the vector $[0,0,1,2,1,0,0,0,0,1]$. 
}
}


\subsection{Knowledge Elicitation}
\label{sec:knowledgeElicitation}

This step focuses on extracting knowledge $K$ 
from teacher models. 
It receives as input a set $T$ of tuples and returns as output the knowledge $K$ generated about these tuples in the form of labels, features, or model predictions. 
DistillER distinguishes the possible approaches according to two criteria: (i) the type of models used as teachers, and (ii) the explicit instructions that are given to the teacher model. We present the methods offered by DistillER for each criterion in the following. 


\subsubsection{Teacher Model}
\label{sec:labeling}

The teacher model is responsible for labelling a given tuple $t_{q,c}$, i.e., for
predicting which candidate $c_i$ in $C$ is the most appropriate match for the query entity $q$. Based on the type of the language model used for labelling, 
we define the following strategies:

\begin{itemize}[leftmargin=*]
    \item \textbf{LLM Annotation}: 
    Each tuple $t_{q,c}$ is serialized into a SELECT prompt, with an LLM selecting one of the candidate entities as the actual match to the query entity. To ensure annotation quality, DistillER leverages medium- or large-sized LLMs, i.e., with at least 8B parameters.
    
    \item \textbf{SLM Annotation}: Unlike LLMs, SLMs are more efficient but lack an interactive interface. They cannot process prompt-based formulations and instead operate only on fixed pairwise inputs $(e_a,e_b)$, which approximate MATCH prompts, while still requiring task-specific training (i.e., fine-tuning) to achieve high effectiveness.
    To that end, DistillER leverages a pre-trained LLM to annotate a portion of the training data as follows: First, it splits the training set into two disjoint parts, $X\%$ and $Y\%$ (with $Y\% = 100\% - X\%$). Next, the LLM annotates the former part, and the resulting noisy labels are used to fine-tune the selected SLM. Finally, the fine-tuned SLM annotates the remaining $Y\%$ of the training set. Note that this may yield more than one match 
    per tuple, but this is in line with this step's assumption about noisy training data. {\color{black} In total, the SLM annotation time can be expressed as:

    \begin{center}
        SLM Annotation time = X\% $\cdot$ LLM Annotation time + SLM Fine-tuning (X\%) + SLM Testing (Y\%).
    \end{center}

    Since the first term dominates the last two, we can safely simplify the above formula as follows:

    \begin{center}
        SLM Annotation time $\approx$ X\% $\cdot$  LLM Annotation time
    \end{center}
    }

    \item \textbf{Multi-Teacher Annotation}: Instead of relying on a single model for annotation, this strategy aggregates the labels generated by a committee of models and annotates each tuple through a simple voting scheme that returns the most frequent label, while in cases of ties a random selection between the most frequent votes is made. Due to the lack of compatibility between LLM and SLM annotation, we combine models from the same family, i.e., all models are LLMs or SLMs. 
\end{itemize}

\vspace{2pt}
\noindent
\textbf{Example.}
\textit{For our example in Figure \ref{fig:prompt_variations} using the LLM Annotator could lead to choosing candidate 2, which is also a Sony Turntable with the same product id. Using the SLM Annotator, it could lead to suggesting candidates 2 and 3, which are both Sony Turntables. Finally, let us assume that we use a committee of three LLMs, which choose as correct the candidates 2, 2, 3, respectively. The Multi-Teacher approach would suggest labeling the candidate 2 as the real match, since it gathers the most votes.}

\subsubsection{Instructions}
This step of the DistillER framework applies only {\color{black} when using LLMs as Teachers.}
Its goal is to determine the form of the Task Description in the SELECT prompts (i.e., the bold part in Figure \ref{fig:prompt_variations}). Two different forms of instructions are supported, differing in the 
format and scope of the output of the SELECT prompt.
DistillER supports the following 
two types of instructions:

\begin{itemize}[leftmargin=*]
    \item \textbf{Answer-only:} The LLM is instructed to identify the most~likely matching 
    candidate $c_i$ from the tuple $t_{q,c}$. This minimal format ensures direct supervision without additional reasoning.  
    \item \textbf{Answer + Explanation:} The LLM is instructed to return not only the matching candidate $c_i$, but also a justification for its choice. This extended format allows DistillER to capture both the decision and the reasoning process behind it, offering richer knowledge to transfer during distillation.  
\end{itemize}

\subsection{Distillation Algorithm}
\label{sec:distillation}

This component focuses on transferring the elicited knowledge $K$, together with the original collection $C$ of tuples, to the student models. It receives as input the training set generated by Data Selection and the Knowledge accumulated by the Knowledge Elicitation step, while its output comprises a trained student model. This process involves two key challenges: (i) choosing an appropriate student model, and (ii) defining an effective training strategy. For the former, DistillER supports LLM and SLM student models. For the latter, we elaborate on the strategies supported by DistillER in the following.

\subsubsection{Supervised Fine-Tuning}

This step fine-tunes a pre-trained model using the labeled data derived from Knowledge $K$ in order to boost its performance on a specific task. 
The two families of student models supported by DistillER 
give rise to different fine-tuning processes:

\begin{itemize}[leftmargin=*]
    \item \textbf{LLM}: This approach fine-tunes an LLM with the labeled tuples generated by the teacher model, using a conversational format. The SELECT prompt ensures exactly one positive answer, providing a straightforward end-to-end prediction without further disambiguation.
    \item \textbf{SLM}: Modeled as a binary classification task, the input comprises  
    fixed pairwise inputs $(e_a,e_b)$, which approximate MATCH prompts,
    and the prediction is 0 (No) or 1 (Yes). Given that the pairwise inputs can produce multiple positive predictions (i.e., multiple candidates in a tuple $t_{q,c}$ are classified as matches), this approach also applies one of the two supported disambiguation strategies:
    (i) {\color{black} utilizing the SELECT prompt on LLMs based only on the classified matches in order to obtain a single final answer}, and (ii) applying a clustering algorithm that leverages the 1-1 constraint between the input data sources; for example, Unique Mapping Clustering (UMC) \cite{DBLP:conf/kdd/Lacoste-JulienPDKGG13} sorts the conflicting matches in decreasing similarity score and retains as true matches the top-weighted one for each query entity. {\color{black} To be noted, the SELECT prompt follows the same structure as the one introduced earlier, but differs in the set of involved candidates, which is a reduced subset originating from the blocking phase.}
\end{itemize}


\vspace{2pt}
\noindent
\textbf{Example.}
\textit{The tuple in the example of Figure \ref{fig:prompt_variations} can be used for training an LLM student model as long as it is accompanied by the id of the detected match (ideally, the correct one, i.e., [2]). 
In the case of SLM fine-tuning, the tuple in Figure \ref{fig:prompt_variations} would be transformed in five training instances (i.e., pairs of $(e_a, e_b)$), one for each candidate entity.  
Each prompt is accompanied by a Yes or No answer, based on whether this candidate is labeled as an answer. If this tuple was used for testing purposes, let us assume that both candidates 2 and 3 were labeled as possible answers. The disambiguation step should detect that candidate 3 has a different product number and thus is not a valid answer.}
    
\subsubsection{Reinforcement Learning} 
\label{sec:rl}

RL provides a mechanism to further refine student models by leveraging preference signals derived from the elicited knowledge $K$. 
The model is trained to maximize alignment with these preferences, adapting dynamically beyond what supervised labels alone can provide. {\color{black} In this work, we focus on RL strategies that are applicable only to Large Language Models, as the considered methods rely on capabilities (e.g., instruction following and rich preference modeling) that are not supported by SLM student models. } The optimization process depends on both the formulation of the reward or preference signals (i.e., on the RL strategy) and the initialization of the student~model (i.e., the model state). 

\textbf{Strategies:} We consider two RL approaches that directly exploit preference signals: 

\begin{itemize}[leftmargin=*]
    \item \textbf{Generalized Reward Policy Optimization (GRPO)} \cite{shao2024deepseekmath}: This is a variant of policy optimization that leverages group-based advantage estimation to efficiently update the student model based on multiple responses per prompt. This approach is particularly effective for complex reasoning tasks with LLMs. We use the following reward functions:

    \begin{itemize}
        \item \textbf{Digit Count Reward:} Scores completions based on the number of digits enclosed in square brackets. Formally:
         \begin{equation}
    R_\text{digit}(c) =
        \begin{cases}
        0 & \text{if } br(c) = 0, \\
        \frac{1}{br(c)} & \text{if } br(c) > 1,
        \end{cases}
    \end{equation}
    where $br(c)$ is the number of digits in brackets $[]$ in \textit{completion} $c$ {\color{black}(i.e., the model-generated output).}
        \item \textbf{Length-based Reward:} Scores completions based on their length after stripping whitespace and newlines. Formally:
        \begin{equation}
    R_\text{length}(c) =
        \begin{cases}
        0 & \text{if } |c| = 0, \\
        1 & \text{if } |c| \in \{1,2\}, \\
        \frac{3}{|c|} & \text{if } |c| > 2,
        \end{cases}
    \end{equation}
    where $c$ is the completion after removing leading/trailing whitespace and newlines.
        \item \textbf{Correct Answer Reward:} Scores completions based on whether the predicted answer matches the ground truth. Formally:
        \begin{equation}
    R_\text{answer}(c) =
        \begin{cases}
        0 & \text{if } pr(c) \ne t, \\
        1 & \text{if } pr(c) = t, \\
        \end{cases}
    \end{equation}
    where $pr(c)$ is the predicted answer based on completion $c$ and $t$ the correct answer.
    \end{itemize}  
    
    Finally, the \textbf{total reward} is computed as a weighted sum of the three individual rewards:
    
    \begin{equation}
        R_\text{total}(c) = w_1 \cdot R_\text{digit}(c) + w_2 \cdot R_\text{length}(c) + w_3 \cdot R_\text{answer}(c)
    \end{equation}

    \vspace{2pt}
    \noindent
    \textbf{Example.} \textit{Extending our running example with equal weights:
    \begin{itemize}
        \item "The correct answer is [2]": This response has $R_\text{total}(c) = \frac{1}{3} \cdot 1 + \frac{1}{3} \cdot \frac{3}{25} + \frac{1}{3} \cdot 1 = 0.706$ 
        \item "Answer: [2]": This response has $R_\text{total}(c) = \frac{1}{3} \cdot 1 + \frac{1}{3} \cdot \frac{3}{11} + \frac{1}{3} \cdot 1 = 0.757$ 
        \item "The correct answer is [2], since [1] and [4] are of a different brand.": This response has $R_\text{total}(c) = \frac{1}{3} \cdot \frac{1}{3} + \frac{1}{3} \cdot \frac{3}{70} + \frac{1}{3} \cdot 1 = 0.458$ 
    \end{itemize}
    This illustrates a preference for concise and precise answers without unnecessary references to other options. 
    }
    
    \item \textbf{Direct Preference Optimization (DPO)} \cite{rafailov2023direct}: This is a preference-based method that bypasses explicit reward modeling and aligns the student model directly with pairwise preference data. 

    \vspace{2pt}
    \noindent
    \textbf{Example.}    
    \textit{Following our example, the model will be trained by conversations, where the prompt is the SELECT prompt and will be given also two preferences. The "chosen" will be the accepted one ([2] in the example of Figure \ref{fig:prompt_variations}) and the "rejected" would be any other, such as [1]. With explanations enabled, the method considers both the answer and the justification. }
\end{itemize}

\textbf{Model State:} The initialization of the student model plays a key role in RL performance:

\begin{itemize}[leftmargin=*]
    \item \textbf{Pre-trained:} Initialized from general pre-training, the model relies on emergent abilities to adapt to preference signals.
    \item \textbf{SFT:} Already fine-tuned on the target task, the model can incorporate preferences more rapidly and effectively.
\end{itemize}

\noindent
Compared to SFT, RL enables continual adaptation, allowing the student model to dynamically refine its predictions in response to evolving preference signals derived from the knowledge $K$. This dynamic guidance can come from either noisy human annotations or model-generated preferences, helping the student model to achieve better alignment with the intended behavior.

\section{Evaluation}
\label{sec:evaluation}

\begin{table*}[!t]
\setlength{\tabcolsep}{4.8pt}
\small
\begin{tabular}{l|cc|cc|cc||cc|cc||cc|cc|cc|}
\cline{2-17}
& \multicolumn{6}{|c||}{Products} & \multicolumn{4}{|c||}{Papers} & \multicolumn{6}{|c|}{Movies} \\
\cline{2-17}
& \multicolumn{2}{|c|}{D2} & \multicolumn{2}{|c|}{D3} & \multicolumn{2}{|c||}{D8} & \multicolumn{2}{|c|}{D4} & \multicolumn{2}{|c||}{D9} & \multicolumn{2}{|c|}{D5} & \multicolumn{2}{|c|}{D6} & \multicolumn{2}{|c|}{D7} \\
\hline
\hline
\multicolumn{1}{|l|}{Name} & Buy & Abt & Amz & GPr & Wmt & Amz & ACM & DBLP & DBLP & Scholar & IMDb & TMDb & IMDb & TVDb & TVDb & TMDb \\
\multicolumn{1}{|l|}{|E|} & 1,064 & 1,076 & 1,354 & 2,935 & 2,554 & 22,068 & 2,272 & 2,576 & 2,476 & 61,350 & 5,117 & 6,042 & 5,117 & 5,178 & 5,178 & 6,042 \\
\multicolumn{1}{|l|}{|A|} & 3 & 3 & 4 & 4 & 6 & 6 & 4 & 4 & 4 & 4 & 13 & 30 & 13 & 9 & 9 & 30 \\
\multicolumn{1}{|l|}{|D|} & \multicolumn{2}{c|}{1,064} & \multicolumn{2}{c|}{1,102} & \multicolumn{2}{c||}{853} & \multicolumn{2}{c|}{2,214} & \multicolumn{2}{c||}{2,304} & \multicolumn{2}{c|}{1,965} & \multicolumn{2}{c|}{966} & \multicolumn{2}{c|}{1,035} \\


\hline
\end{tabular}

\caption{The datasets used in the experimental evaluation. $|E|$ stands for the number of entities, $|A|$ for the number attributes and $|D|$ for the number of duplicates. 
}
\label{tab:datasets}
\end{table*}

In this section, we present a comprehensive evaluation of DistillER, examining its performance across multiple stages of the pipeline. 

\subsection{Experimental Setup}
\label{subsec:setup_results}
\noindent \textbf{Datasets.} We evaluate our approach on eight real-world entity resolution datasets widely used in prior work \cite{DBLP:conf/sigmod/MudgalLRDPKDAR18,DBLP:journals/corr/abs-2405-16884}. Each entity is represented as a set of \texttt{<attribute, value>} pairs, which we serialize into a single textual description. This representation is flexible and applies to both tabular and semi-structured data sources. Table \ref{tab:datasets} summarizes the dataset statistics. For training, we sample 10\% of the entities from each dataset to construct a \textit{global training set}.

\noindent \textbf{Models.} We used four open-source LLMs: Llama-3.1:8b, Llama-3.1:70b, Qwen-2.5:14b and Qwen-2.5:32b\footnote{https://ollama.com/library/\{X\} llama3.1:8b, llama3.1:70b, qwen2.5:14b qwen2.5:32b} and  2 SLMs: S-MiniLM\footnote{https://huggingface.co/sentence-transformers/all-MiniLM-L6-v2} and RoBERTa\footnote{https://huggingface.co/FacebookAI/roberta-base}.

\noindent \textbf{Evaluation Metrics.} We report the F-Measure (F1) across all experiments. {\color{black} Note that for models utilizing SELECT prompts, 
precision=recall, 
but for MATCH prompts 
precision $\neq$ recall.}

\noindent \textbf{Settings.} For the pre-trained LLMs, we used models provided by Ollama\footnote{https://ollama.com/}, while fine-tuning is performed with Unsloth\footnote{https://unsloth.ai/}. All experiments were executed on a server with Ubuntu 20.04, AMD Ryzen Threadripper 3960X 24-Core processor, 256 GB RAM and an RTX 4090 GPU. 

\noindent \textbf{Blocking.} For blocking, we adopt a standard approach in literature \cite{DBLP:journals/pvldb/ZeakisPSK23, zeakis2025depth}: entities are serialized with S-GTR-T5, candidate pairs are refined via Reciprocal Pruning, and FAISS is used for indexing and retrieving the top-$N$ candidates per entity.


\begin{table*}[]
\small
\setlength{\tabcolsep}{4.6pt}
\begin{tabular}{|cc|cc|cc|cc||cc|cc||cc|cc|cc|}
\cline{3-18}
\multicolumn{2}{c|}{} & \multicolumn{2}{c|}{D2} & \multicolumn{2}{c|}{D3} & \multicolumn{2}{c||}{D8} & \multicolumn{2}{c|}{D4} & \multicolumn{2}{c||}{D9} & \multicolumn{2}{c|}{D5} & \multicolumn{2}{c|}{D6} & \multicolumn{2}{c|}{D7} \\ 
 \multicolumn{2}{l|}{} & Buy & Abt & Amz & GPr & Wmt & Amz & ACM & DBLP & DBLP & Scholar & IMDb & TMDb & IMDb & TVDb & TVDb & TMDb \\
 \hline
 \hline
\multirow{2}{*}{$|e_i|$} & 50\% & \multicolumn{1}{c}{145} & 370 & \multicolumn{1}{c}{579} & 368 & \multicolumn{1}{c}{233} & 235 & \multicolumn{1}{c}{239} & 213 & \multicolumn{1}{c}{213} & 193 & \multicolumn{1}{c}{154} & 56 & \multicolumn{1}{c}{154} & 80 & \multicolumn{1}{c}{80} & 56 \\ 
 & 95\% & \multicolumn{1}{c}{257} & 528 & \multicolumn{1}{c}{6,894} & 432 & \multicolumn{1}{c}{284} & 301 & \multicolumn{1}{c}{315} & 284 & \multicolumn{1}{c}{281} & 266 & \multicolumn{1}{c}{217} & 682 & \multicolumn{1}{c}{217} & 715 & \multicolumn{1}{c}{715} & 682 \\ \hline
\multirow{2}{*}{$|C_i|$} & 50\% & \multicolumn{2}{c|}{6} & \multicolumn{2}{c|}{7} & \multicolumn{2}{c||}{10} & \multicolumn{2}{c|}{4} & \multicolumn{2}{c||}{7} & \multicolumn{2}{c|}{3} & \multicolumn{2}{c|}{2} & \multicolumn{2}{c|}{5} \\ 
 & 95\% & \multicolumn{2}{c|}{10} & \multicolumn{2}{c|}{10} & \multicolumn{2}{c||}{10} & \multicolumn{2}{c|}{9} & \multicolumn{2}{c||}{10} & \multicolumn{2}{c|}{10} & \multicolumn{2}{c|}{10} & \multicolumn{2}{c|}{10} \\ \hline
$|EP_i|$ & 50\% & \multicolumn{2}{c|}{2,362} & \multicolumn{2}{c|}{3,155} & \multicolumn{2}{c||}{2,583} & \multicolumn{2}{c|}{1,091} & \multicolumn{2}{c||}{1,564} & \multicolumn{2}{c|}{322} & \multicolumn{2}{c|}{314} & \multicolumn{2}{c|}{360} \\ \hline
\multirow{2}{*}{$|OP_i|$} & 50\% & \multicolumn{2}{c|}{3,408} & \multicolumn{2}{c|}{3,413} & \multicolumn{2}{c||}{3,191} & \multicolumn{2}{c|}{2,143} & \multicolumn{2}{c||}{2,345} & \multicolumn{2}{c|}{1,633} & \multicolumn{2}{c|}{1,440} & \multicolumn{2}{c|}{3,619} \\ 
 & 95\% & \multicolumn{2}{c|}{5,012} & \multicolumn{2}{c|}{5,563} & \multicolumn{2}{c||}{3,853} & \multicolumn{2}{c|}{3,186} & \multicolumn{2}{c||}{3,123} & \multicolumn{2}{c|}{2,501} & \multicolumn{2}{c|}{2,154} & \multicolumn{2}{c|}{12,413} \\ \hline

\multicolumn{2}{|c|}{|P|} & \multicolumn{2}{|c|}{0.95} & \multicolumn{2}{c|}{0.71} & \multicolumn{2}{|c||}{0.36} & \multicolumn{2}{c|}{0.97} & \multicolumn{2}{|c||}{0.83} & \multicolumn{2}{|c|}{0.31} & \multicolumn{2}{|c|}{0.14} & \multicolumn{2}{c|}{0.19} \\ \hline
\end{tabular}

\caption{{\color{black} Prompt-related statistics per dataset. $|e_i|$ stands for
the serialized entity length in characters, $|C_i|$ for
the blocking candidate list size, $|EP_i|$ and $|OP_i|$ 
for the estimated 
and the observed prompt lengths
in characters and $|P|$ for the positive rate, {\color{black} showing 50\% (median) and 90\% percentiles of each statistic across all entity pairs}.
}}
\label{tab:prompt_stats}

\end{table*}

\subsection{Data Selection}
{\color{black} \noindent \textbf{Q1: Which is the most effective strategy for Data Selection?}}







\begin{table}
\centering

\begin{tabular}{|c|c|rr|rc|c|}
\cline{3-6}

\multicolumn{2}{c|}{} & \multicolumn{2}{|c|}{Ranking} & \multicolumn{2}{|c|}{Clustering} & \multicolumn{1}{c}{} \\

\cline{2-7}
\multicolumn{1}{c|}{} & Random & Max & Top-2 & Agg & KMeans & Sampled \\
\hline
\hline
D2 & 0.96 & 0.94 & 0.95 & 0.99 & 0.94 & 0.75 \\
D3 & 0.75 & 0.58 & 0.61 & 0.72 & 0.72 & 0.75 \\
D4 & 0.98 & 0.90 & 0.93 & 0.98 & 0.99 & 0.75 \\
D5 & 0.42 & 0.74 & 0.67 & 0.44 & 0.44 & 0.75 \\
D6 & 0.17 & 0.46 & 0.44 & 0.17 & 0.14 & 0.75 \\
D7 & 0.20 & 0.43 & 0.42 & 0.23 & 0.24 & 0.75 \\
D8 & 0.34 & 0.50 & 0.46 & 0.38 & 0.35 & 0.75 \\
D9 & 0.94 & 0.82 & 0.83 & 0.91 & 0.88 & 0.75 \\
\hline
Mean & 0.59 & 0.67 & 0.67 & 0.60 & 0.59 & 0.75 \\
\hline
\end{tabular}
\caption{{\color{black} Positive Ratio on different data selection strategies. }}
\label{tab:sel_distr}
\end{table}

\begin{table*}
\centering

\begin{tabular}{|c|rr|rr|rr|rr|rr|rr|}
\cline{2-13}

\multicolumn{1}{c|}{} & \multicolumn{2}{c|}{} & \multicolumn{4}{|c|}{Ranking} & \multicolumn{4}{|c|}{Clustering} & \multicolumn{2}{c|}{} \\

\multicolumn{1}{c|}{} & \multicolumn{2}{|c|}{Random} & \multicolumn{2}{|c|}{Max} & \multicolumn{2}{|c|}{Top-2} & \multicolumn{2}{|c|}{Agg} & \multicolumn{2}{|c|}{KMeans}  & \multicolumn{2}{|c|}{Sampled} \\

\cline{2-13}
\multicolumn{1}{c|}{} & L:8b & Q:14b & L:8b & Q:14b & L:8b & Q:14b & L:8b & Q:14b & L:8b & Q:14b & L:8b & Q:14b\\
\hline
\hline

D2 & 0.89 & 0.94 & 0.92 & 0.92 & 0.92 & 0.91 & 0.88 & 0.90 & 0.87 & 0.90 & 0.68 & 0.86 \\
D3 & 0.58 & 0.69 & 0.57 & 0.64 & 0.54 & 0.61 & 0.46 & 0.59 & 0.51 & 0.58 & 0.61 & 0.69 \\
D4 & 0.96 & 0.97 & 0.86 & 0.93 & 0.88 & 0.95 & 0.96 & 0.97 & 0.96 & 0.99 & 0.74 & 0.81 \\
D5 & 0.61 & 0.78 & 0.87 & 0.98 & 0.81 & 0.94 & 0.69 & 0.86 & 0.67 & 0.83 & 0.85 & 0.94 \\
D6 & 0.43 & 0.81 & 0.61 & 0.72 & 0.57 & 0.71 & 0.44 & 0.83 & 0.39 & 0.80 & 0.77 & 0.82 \\
D7 & 0.34 & 0.66 & 0.50 & 0.63 & 0.48 & 0.63 & 0.36 & 0.70 & 0.40 & 0.70 & 0.74 & 0.88 \\
D8 & 0.36 & 0.67 & 0.59 & 0.77 & 0.55 & 0.75 & 0.44 & 0.65 & 0.38 & 0.65 & 0.64 & 0.82 \\
D9 & 0.90 & 0.96 & 0.83 & 0.91 & 0.82 & 0.88 & 0.85 & 0.93 & 0.83 & 0.92 & 0.73 & 0.87 \\
\hline
Mean & 0.63 & 0.81 & 0.72 & 0.81 & 0.70 & 0.80 & 0.63 & 0.80 & 0.63 & 0.80 & 0.72 & 0.84 \\
\hline
\end{tabular}

\caption{{\color{black} F1-score on training data for different data selection strategies. L:8b stands for Llama-3.1:8b and Q:14b for Qwen-2.5:14b.}}
\label{tab:sel_effectiveness}
\end{table*}

    

In this experiment, we evaluate unsupervised strategies for selecting high-quality tuples for training without relying on labels. From each dataset, we sample 10\% of the tuples, resulting in a global training set of 2,181 tuples.
To approximate the 3-to-1 positive-to-negative ratio suggested in \cite{zeakis2025avenger,DBLP:journals/corr/abs-2405-16884}, we set $p=7.5\%$ and $n=2.5\%$ for both the Ranking-based methods (Max and Top-2) and the Clustering-based methods (KMeans and Agglomerative Clustering). As a purely unsupervised baseline, we include Random selection of 10\% of the tuples. Additionally, using available labels, we construct a second baseline (Sampled) by explicitly selecting 7.5\% positive and 2.5\% negative tuples. 
{\color{black} To evaluate each strategy, we employ two LLMs, Llama-3.1:8b and Qwen-2.5:14b, which are used in their pre-trained form to generate predictions over the different data splits constructed by each selection method. We compare their performance across these splits, assessing their sensitivity to dataset skewness and 
the importance of constructing well-balanced training subsets. 
}


{\color{black}To this end, we define the Positive Ratio as ``positive instances'' / ``total instances'', with values larger than 0.5 indicating the prevalence of positive pairs over the negative ones. The Positive Ratio is reported in Table~\ref{tab:sel_distr}, which highlights } the strong class imbalance in the generated datasets. Random selection disproportionately favors the negative class, as very few query tuples are indeed positive (see also Table~\ref{tab:datasets}). Ranking approximates the 3-to-1 ratio in some datasets (e.g., D5) under both Max and Top-2, but collapses to a 50\% split in others (e.g., D6 and D8). Nonetheless, Ranking generally preserves a stronger representation of positives compared to Random or Clustering. In contrast, Clustering often favors the negative class (e.g., D6, D8), which degrades downstream performance. {\color{black} This behavior is highlighted in datasets D5–D8, which are negative-dominant and where entities contain multiple attributes and tokens, leading to higher chances of false-positive token overlaps and wrong similarity scores among all candidates in the tuple. Consequently, the resulting similarity vectors may be incorrectly assigned to the negative cluster, preventing most methods from achieving a valid class separation.}

The results in Table~\ref{tab:sel_effectiveness} confirm these observations. For Llama-3.1:8b, highly imbalanced samples (Random, Clustering) lead to poor performance, while Ranking matches the labeled Sampled baseline under both Max and Top-2. Qwen-2.5:14b, however, being larger, proves less sensitive to the selection strategy, consistently yielding higher scores across all methods. 

{\color{black} \textit{Summary: Ranking emerges as the most reliable unsupervised selection method, maintaining a balanced class distribution and achieving comparable performance to the labeled Sampled baseline under both Max and Top-2 strategies.} }


{\color{black}
\subsection{Teacher Comparison}
\noindent \textbf{Q2: Which are the most effective and efficient Teachers?}
}

In this experiment, we compare the effectiveness and efficiency of different {\color{black}teachers in generating single-answers {\color{black} when annotating tuples.}}
For effectiveness, we evaluate F1 on the training data, while for efficiency, we measure the required time. We consider two categories of {\color{black}teacher models}, LLMs and SLMs, which are studied separately and compared in the final paragraph.

\textbf{LLM {\color{black}Teachers}}: In this experiment, we use all four LLMs and the multi-teacher setting, comparing them by effectiveness. The results are shown in Table \ref{tab:annot_llm_effectiveness}.

\begin{table}
\centering

\begin{tabular}{|c|rr|rr|c|}
\cline{2-6}
\multicolumn{1}{c|}{} & \multicolumn{2}{|c|}{Llama-3.1} & \multicolumn{2}{|c|}{Qwen-2.5} & \multirow{2}{*}{Multi-LLM}\\
\cline{2-5}
\multicolumn{1}{c|}{} & 8b & 70b & 14b & 32b & \\
\hline
\hline
D2 & 0.92 & \textbf{0.97} & 0.92 & 0.92 & 0.95 \\
D3 & 0.57 & 0.66 & 0.64 & 0.65 & \textbf{0.67} \\
D4 & 0.86 & 0.94 & 0.93 & \textbf{0.96} & 0.93 \\
D5 & 0.87 & 0.98 & 0.98 & \textbf{0.99} & 0.98 \\
D6 & 0.61 & 0.71 & 0.72 & \textbf{0.73} & \textbf{0.73} \\
D7 & 0.50 & \textbf{0.66} & 0.63 & 0.63 & 0.65 \\
D8 & 0.59 & 0.74 & 0.77 & \textbf{0.84} & 0.82 \\
D9 & 0.83 & 0.89 & 0.91 & \textbf{0.96} & 0.93 \\
\hline
Mean & 0.72 & 0.82 & 0.81 & \textbf{0.83} & \textbf{0.83} \\
\hline
\end{tabular}

\caption{F1-score {\color{black}of the annotated} training data per LLM teacher.}
\label{tab:annot_llm_effectiveness}
\end{table}

Within the Llama-3.1 family, we observe a clear improvement when using the larger model, increasing F1 from 0.72 to 0.82. {\color{black} Nonetheless, both models seem to underperform on dataset D3, D7 and D8, which is based on their their excessively large entity profiles. Examining the statistics in Table \ref{tab:prompt_stats} and the entity size $|e_i|$ in characters, we confirm that D3 has the larger entities, ranging from 579 on average to 7K characters. Combined with the blocking candidate list size $|C_i|$ - five candidates per entity after blocking for most datasets and ten for D8 - we expect larger prompts $|EP_i|$ in the product datasets. Beside the product datasets, in practice we notice that D7 also has larg prompt size on average ($|OP_i|$).}
Similarly, in the Qwen-2.5 family, the larger model slightly improves performance, from 0.81 to 0.83. Overall, the top-2 performances are achieved by the larger models. {\color{black} To measure the complementarity of the models, we compare the average recall of the individual Llama and Qwen models (0.82) with the recall obtained by combining all predictions from all models (0.88). This indicates that the models are complementary {\color{black}to some extent}. Accordingly, by applying a simple majority-voting scheme, as in \cite{zeakis2025avenger}, Multi-LLM achieves an average F1 of 0.83; however, it requires running four teacher models in parallel. As a result, this approach does not justify the effectiveness–efficiency trade-off and can be reasonably omitted.}



\textit{Summary: Using LLMs as teachers can produce high-quality labels that are close to the ground-truth data, whereas the multi-teacher setting does not justify the effectiveness–efficiency tradeoff.}

\textbf{SLM Teachers}: First, we compare the two SLM teachers trained on different noisy labels generated by the best LLM teachers. We use {\color{black} 20\% of the training data (8\% of the total data) is annotated by an LLM and is then used to fine-tune an SLM, which then annotates the remaining 80\% (8\% of the total data).} 
The results are shown in Table \ref{tab:annot_slm_llm}. 

\begin{table}
\centering

\begin{tabular}{|c|c|cc||c|cc|}
\cline{2-7}
\multicolumn{1}{c|}{} & \multicolumn{3}{|c||}{S-MiniLM} & \multicolumn{3}{|c|}{RoBERTa} \\
\cline{2-7}
\multicolumn{1}{c|}{} & GT & L:70b & Q:32b & GT & L:70b & Q:32b\\
\hline
\hline

D2 & 0.48 & 0.48 & 0.41 & 0.58 & 0.51 & \textbf{0.61} \\
D3 & 0.38 & 0.37 & 0.43 & 0.48 & \textbf{0.49} & 0.46 \\
D4 & 0.78 & 0.77 & 0.75 & \textbf{0.79} & 0.76 & 0.78 \\
D5 & 0.84 & 0.85 & 0.84 & \textbf{0.86} & 0.85 & 0.85 \\
D6 & 0.71 & 0.65 & 0.66 & \textbf{0.73} & 0.65 & 0.66 \\
D7 & 0.68 & 0.66 & 0.62 & \textbf{0.75} & 0.66 & 0.64 \\
D8 & 0.69 & 0.63 & 0.64 & \textbf{0.72} & 0.68 & \textbf{0.72} \\
D9 & 0.76 & 0.76 & 0.76 & \textbf{0.80} & \textbf{0.80} & 0.79 \\
\hline
Mean & 0.67 & 0.65 & 0.64 & \textbf{0.71} & 0.68 & 0.69 \\

\hline
\end{tabular}

\caption{F1-score {\color{black}of the annotated} training data for different SLM teachers, when trained on noisy data from different LLM teachers and ground-truth labels (20\%). L stands from Llama-3.1 and Q for Qwen-2.5}
\label{tab:annot_slm_llm}
\end{table}

\begin{figure}[!t]
\centering
\includegraphics[trim=0 0 0 0, width=0.8\linewidth]{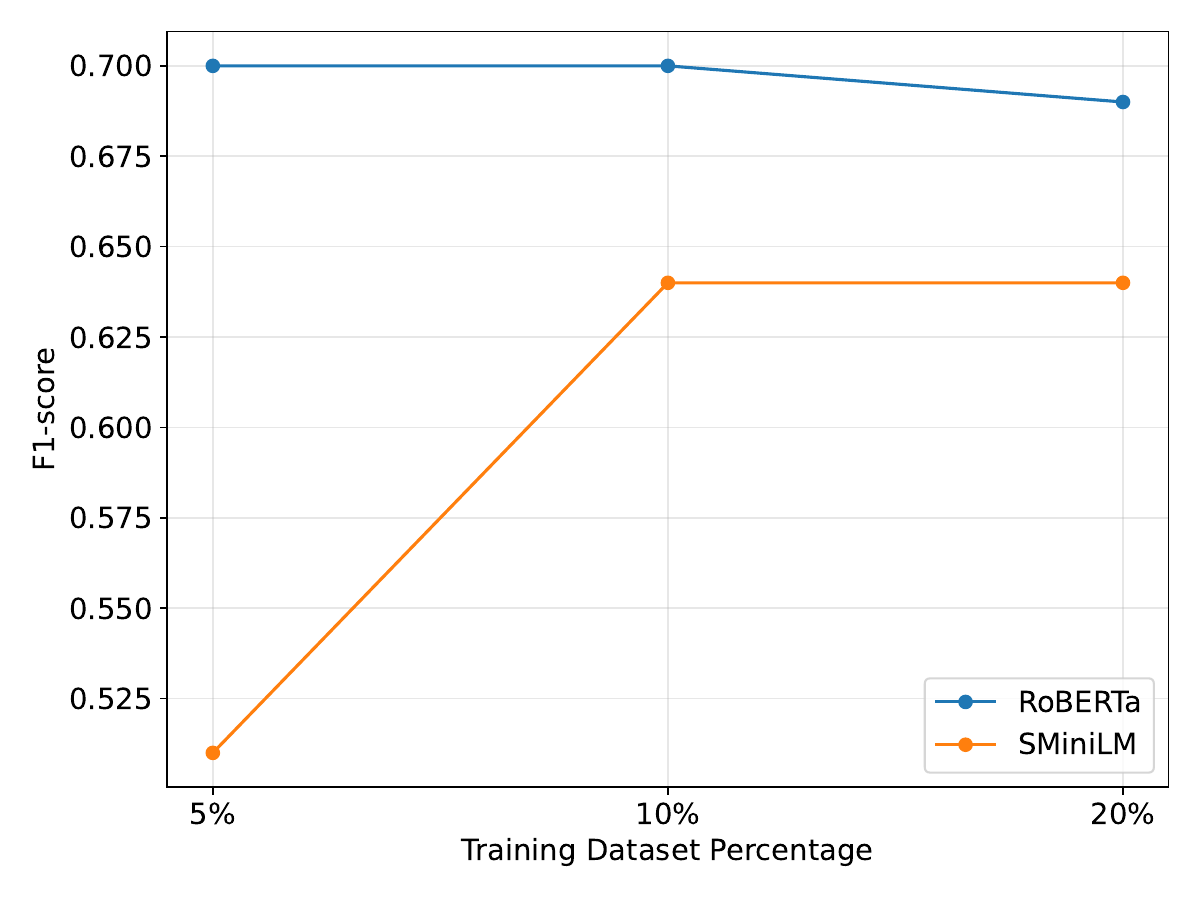}
\vspace{-5pt}
\caption{Average f1-score on training data for different SLM teachers using Qwen-2.5:32b labels, when trained on different training sizes.}
\label{fig:annot_slm_size}
\vspace{-5pt}
\end{figure}

S-MiniLM shows the poorest average performance (0.65), whereas RoBERTa performs better, reaching 0.69 on average with noisy labels from Qwen-2.5:32b. {\color{black} This should be attributed to RoBERTa being a larger and more robust model {\color{black}(125M vs 22M parameters), while S-MiniLM emphasizes time efficiency}. Nonetheless, the worse performance is observed for all teachers in product datasets, which indicates that product representations include common tokens that might disorientate the {\color{black}teacher} model in detecting true matches. For example, we could have an Ethernet cable described as "Cat6 Ethernet cable 5m" while another entry appears as "Cat6 Ethernet cable 15m," where the shared tokens overwhelm the subtle but crucial difference in length. In addition, }when trained on ground-truth data, both SLM teachers perform slightly better, though the difference is small (less than 5\%).

Next, we evaluate the robustness of SLM models when varying the size of the training data. We reduce the original 20\% to 10\% and 5\%, using only noisy labels from Qwen-2.5:32b. The results are shown in Figure \ref{fig:annot_slm_size}. 

As expected, S-MiniLM is most affected, dropping from 0.64 to 0.50, a 20\% decline. {\color{black} RoBERTa, originally trained on more and larger datasets and  designed to be more robust, shows a more stable performance across all dataset percentages.}

\textit{Summary: Using SLM teachers can produce good-quality labels, especially when using RoBERTa, {\color{black} which is quite effective and robust, even when fine-tuned on very few training data. }}


{\color{black}
\textbf{Comparison}: We compare all teachers from each category with regard to effectiveness (i.e., average F1-score) and time efficiency (i.e., total time). The results are reported in Figure \ref{fig:annot_comparison}. 
{\color{black}Note that total time for LLM teachers includes the time required to annotate all training data, while for SLM teachers, see the respective formula in Section \ref{sec:knowledgeElicitation}.}

\begin{figure}[!t]
\centering
\includegraphics[trim=0 0 0 0, width=1.0\linewidth]{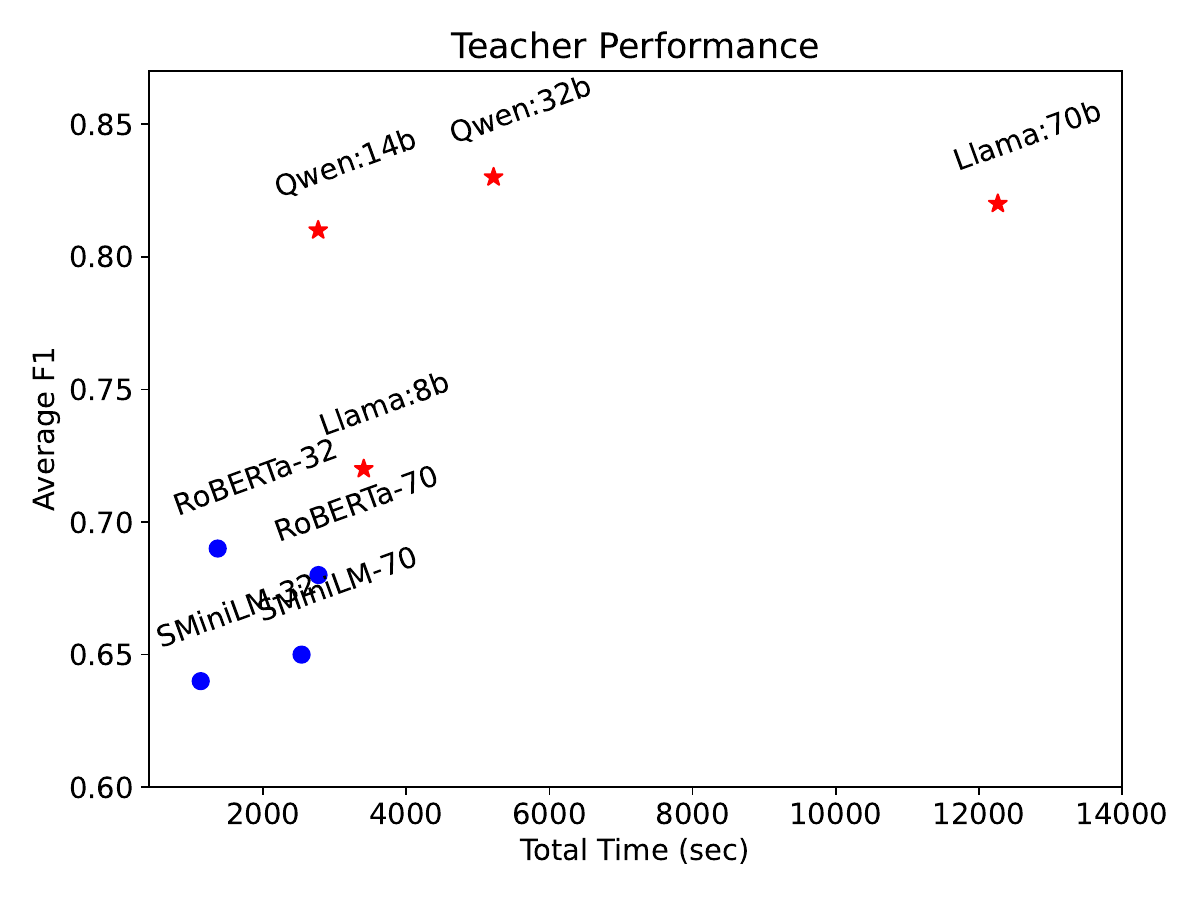}
\caption{{\color{black} Trade-off between effectiveness and time efficiency across different teachers.}}
\label{fig:annot_comparison}
\end{figure}

LLM teachers clearly outperform SLMs in effectiveness, achieving average F1 scores above 0.80, while SLM teachers do not exceed 0.70. In terms of time efficiency, SLM teachers require 
{\color{black} $\approx20\%$} of the {\color{black}respective LLM teachers} inference time. {\color{black}The reason is that LLM inference dominates the run-time of SLM teachers, but applies to just 20\% of the training data.} 
Overall, 
larger models (Qwen:32b, Llama:70b) 
confirm the expectation of higher effectiveness accompanied by higher inference times, whereas 
smaller models result in lower inference times but reduced effectiveness. This 
{\color{black}justifies our goal} to combine the efficiency of smaller models with the effectiveness of larger models.

The following experiments exclusively consider 
Qwen:32b and RoBERTa trained on Qwen:32b labels as LLM and SLM teachers.

}
\textit{Summary: Regarding effectiveness, LLM teachers are clearly superior but require more time, while SLM teachers offer a good tradeoff with much faster execution.}

\subsection{Supervised Fine-Tuning} 
\label{sec:eval_sft}

\noindent \textbf{Q3: How do fine-tuned models on noisy labels perform?}

\begin{table}
\centering

\begin{tabular}{|c|rcc||r|rr|}
\cline{5-7}
\multicolumn{4}{c|}{} & \multicolumn{3}{c|}{Fine-Tuning} \\
\cline{2-7}
\multicolumn{1}{c|}{} & UMC & Student & Teacher & GT & LLM & SLM\\
\hline
\hline
D2 & 0.79 & 0.88 & \textbf{0.92} & 0.75 & 0.88 & 0.48 \\
D3 & 0.49 & 0.56 & \textbf{0.65} & 0.51 & \textbf{0.65} & 0.41 \\
D4 & 0.98 & 0.96 & \textbf{0.99} & \textbf{0.99} & \textbf{0.99} & 0.85 \\
D5 & 0.40 & 0.63 & 0.84 & 0.85 & \textbf{0.91} & 0.89 \\
D6 & 0.17 & 0.49 & 0.89 & 0.89 & \textbf{0.90} & \textbf{0.90} \\
D7 & 0.19 & 0.34 & 0.77 & \textbf{0.81} & 0.77 & 0.79 \\
D8 & 0.20 & 0.33 & \textbf{0.83} & \textbf{0.83} & 0.82 & 0.79 \\
D9 & 0.89 & 0.86 & \textbf{0.92} & 0.90 & 0.88 & 0.65 \\
\hline
Mean & 0.51 & 0.63 & \textbf{0.85} & 0.82 & \textbf{0.85} & 0.72 \\
\hline
\end{tabular}

\caption{ {\color{black} F1-score for fine-tuned Llama-3.1:8b on testing data for different teachers compared to the main relevant baselines. LLM is Qwen-2.5:32b and SLM is RoBERTa on Qwen-2.5:32b labels.}}
\label{tab:ft_llm}
\end{table}

In this experiment, we study the performance of models fine-tuned on noisy labels. Based on the underlying model, we distinguish two categories: LLMs and SLMs.

{\color{black} For LLM training, we use LoRA with rank 16, $\alpha=16$, and no dropout for efficient adaptation, reducing memory usage compared to full fine-tuning.} Training is conducted for three epochs with a learning rate of $2\cdot10^{-4}$, gradient accumulation steps of 4, and the AdamW 8-bit optimizer. We also apply 4-bit quantization to reduce memory usage. For SLM training, we follow the EMTransformer approach \cite{brunner2020entity}.

\textbf{Fine-tuned LLMs}: We use the best teachers of each category from the previous section
{\color{black} and we fine-tune Llama-3.1:8b using these noisy labels to compare it against: (i) Unique Mapping Clustering (UMC), a baseline algorithm for unsupervised Entity Matching, (ii) its pre-trained version (Student), (iii) the {\color{black} pre-trained version of the LLM Teacher (Qwen:32b) }
(iv) its fine-tuned version on on ground-truth data (GT). The results are shown in Table \ref{tab:ft_llm}. }

The {\color{black} student trained from the LLM teacher} consistently outperforms the corresponding {\color{black} student trained from the SLM one}, especially in D2 and D3, which are product datasets where labels are 
problematic, as shown in Table \ref{tab:annot_slm_llm}. Both {\color{black}students} surpass the UMC baseline and the pre-trained model, confirming that fine-tuning improves performance even with noisy labels. Interestingly, the {\color{black} student trained from the LLM teacher} surpasses the model trained on ground-truth labels, achieving an average F1 of 0.85. This can be explained by the fact that annotations involving harder or borderline cases provide stronger training opportunities. Exposure to these cases during fine-tuning helps the model to better distinguish patterns, improving overall performance. {\color{black} This also applies to the Teacher, since in some cases, especially D5, the {\color{black} student trained from the LLM teacher} surpasses the Teacher. Overall, the {\color{black} student trained from the LLM teacher} maintains the same high effectiveness of the Teacher, which is one of the goals of the Knowledge Distillation.}


\textit{Summary: Fine-tuning LLMs on noisy labels generated by LLMs surpasses baselines as well as fine-tuning on SLM-generated labels.}

\textbf{Fine-tuned SLMs}: We now evaluate fine-tuned SLMs using S-MiniLM and RoBERTa, trained on noisy labels from Qwen-2.5:32b and RoBERTa. Baselines include SLMs fine-tuned on ground-truth labels and the UMC algorithm. The results appear in Table~\ref{tab:ft_slm}.

\begin{table}
\centering

\begin{tabular}{|r|r||r|rc||r|rc|}
\cline{3-8}

\multicolumn{2}{c|}{} & \multicolumn{3}{|c||}{S-MiniLM} & \multicolumn{3}{|c|}{RoBERTa} \\

\cline{2-8}
\multicolumn{1}{c|}{} & UMC & GT & LLM & SLM & GT & LLM & SLM \\
\hline
\hline

D2 & \textbf{0.79} & 0.65 & 0.65 & 0.47 & 0.78 & 0.75 & 0.51 \\
D3 & \textbf{0.49} & 0.31 & 0.31 & 0.15 & 0.40 & 0.41 & 0.15 \\
D4 & \textbf{0.98} & 0.96 & 0.96 & 0.92 & 0.97 & \textbf{0.98} & 0.88 \\
D5 & 0.40 & 0.73 & 0.71 & 0.73 & 0.79 & \textbf{0.81} & 0.73 \\
D6 & 0.17 & 0.37 & 0.40 & 0.48 & 0.53 & \textbf{0.55} & 0.42 \\
D7 & 0.19 & 0.60 & 0.57 & \textbf{0.62} & 0.49 & 0.61 & 0.56 \\
D8 & 0.20 & 0.42 & 0.45 & 0.29 & 0.44 & \textbf{0.57} & 0.28 \\
D9 & \textbf{0.89} & 0.86 & 0.86 & 0.76 & \textbf{0.89} & 0.86 & 0.72 \\
\hline
Mean & 0.51 & 0.61 & 0.61 & 0.55 & 0.66 & \textbf{0.69} & 0.53 \\
\hline
\end{tabular}

\caption{F1-score for fine-tuned SLMs on testing data for different teachers compared with baselines. LLM is Qwen-2.5:32b and SLM is RoBERTa on Qwen-2.5:32b labels.}
\label{tab:ft_slm}
\end{table}

Within each SLM, the {\color{black}SLM-based annotations underperform the LLM-based ones by 10\% and 23\%, on average, for S-MiniLM and RoBERTa, respectively.}
Among the two, RoBERTa is more effective, reaching an average F1 of 0.69 {\color{black}when using the LLM-based annotations
while S-MiniLM is reduced to 0.61. This}
can be attributed to RoBERTa being a larger and more robust model. Both models perform comparably to the versions trained on ground-truth data, with RoBERTa 
surpassing by 3\%.
All models outperform the UMC baseline, which exhibits poor performance.

\textit{Summary: Fine-tuning SLMs on noisy labels generated by LLMs surpasses baselines as well as fine-tuning on SLM-generated labels.}





{\color{black}
\textbf{Comparison}: We now compare all fine-tuned models from each category with regard to effectiveness (average f1) and efficiency (inference time). The effectiveness results are reported in Tables \ref{tab:ft_llm} \& \ref{tab:ft_slm} and the efficiency ones in Table \ref{tab:ft_time_comparison}.





\begin{table}
\small
\setlength{\tabcolsep}{3.5pt}
\begin{tabular}{|c|cc||rr||rr|rr|}
\cline{4-9}
\multicolumn{3}{c|}{} & \multicolumn{2}{c||}{\multirow{2}{*}{FT LLM}} & \multicolumn{4}{c|}{FT SLM} \\
\multicolumn{3}{c|}{} & \multicolumn{2}{c||}{} & \multicolumn{2}{c|}{RoBERTa} & \multicolumn{2}{c|}{SMiniLM} \\
\cline{2-9}
\multicolumn{1}{c|}{}& Stud. & Teac. & LLM & SLM & LLM & SLM & LLM & SLM \\
\hline
\hline
Ann. & 0 & 0 & 5,219 & 1,364 & 5,219 & 1,364 & 5,219 & 1,364 \\
Train & 0 & 0 & 4,827 & 4,793 & 1,465 & 1,514 & 378 & 393 \\
\hline
D2 & 1,495 & 2,801 & 2,145 & 2,057 & 12 & 12 & 3 & 3 \\
D3 & 2,151 & 3,329 & 2,771 & 2,700 & 25 & 25 & 7 & 6 \\
D4 & 2,118 & 4,663 & 4,560 & 4,414 & 19 & 19 & 5 & 5 \\
D5 & 7,099 & 7,135 & 7,685 & 7,438 & 42 & 42 & 11 & 11 \\
D6 & 8,446 & 6,628 & 7,123 & 6,869 & 31 & 31 & 8 & 8 \\
D7 & 7,562 & 8,264 & 9,873 & 9,582 & 70 & 70 & 18 & 18 \\
D8 & 10,306 & 5,898 & 5,148 & 4,911 & 55 & 55 & 15 & 15 \\
D9 & 3,111 & 5,473 & 5,288 & 4,795 & 26 & 26 & 7 & 7 \\
\hline
Sum & 42,291 & 44,195 & 54,644 & 48,926 & 6,969 & 3,163 & 5,675 & 1,835 \\
\hline
\end{tabular}

\caption{{\color{black} Times (seconds) for Supervised Fine-Tuning on Llama-3.1:8b on testing data for different teachers compared with baselines. LLM is Qwen-2.5:32b and SLM is RoBERTa on Qwen-2.5:32b labels.}}
\label{tab:ft_time_comparison}
\end{table}


The fine-tuned LLMs and SLMs have a clean distinction in their inference times, with the latter being faster by two orders of magnitude. This should be attributed to the much smaller size of SLMs than LLMs. For the same reason, the training times of SMLs is faster than LLMs by up to a whole order of magnitude. 
Within the SLM category, SMiniLM, being smaller than RoBERTa, reduces the training time by 3/4.

Considering their overall performance when compared to the Teacher, the SLMs have lower effectiveness, meaning that even with their fast training, they cannot adapt and reach the peak performance of the Teacher. LLMs on the other hand can improve, since we clearly see an increase from the average 0.63 to 0.72 with the SLM Teacher and to 0.85 with the LLM Teacher. Notably, the last one is the same average performance with that of the Teacher's. Comparing their inference times, we notice similar behavior, which can be explained by the fact that the two models are not that different in sizes (Student $\approx$ 8b, Teacher $\approx$ 32b), nonetheless, they have different requirements in GPU resources (Student $\approx$ 5GB, Teacher $\approx$ 20GB).

\textit{Summary: The fine-tuned LLM models require more time than their SLM counterparts, but achieve similar performance in effectiveness with the Teacher, {\color{black} while shrinking the memory requirements.}
}
}

\subsection{Reinforcement-Learning}

\noindent \textbf{Q4: How does Reinforcement Learning on noisy labels affect model performance?}

In this experiment, we study the effect of RL on noisy labels across two dimensions:
the choice of teachers and labels and the version of the underlying model (pre-trained or fine-tuned).

{\color{black} 
{\color{black} For training, we use LoRA with rank 8, $\alpha=16$, and no dropout for efficient adaptation, reducing memory usage compared to full fine-tuning}. Training is conducted for one epoch (by default) with a learning rate of $2\cdot10^{-4}$, gradient accumulation steps of 4, and the AdamW 8-bit optimizer. We also apply 4-bit quantization to reduce memory usage.
}

\textbf{Teachers}: We evaluate the impact of different teachers on both effectiveness and efficiency. We select the best teachers of each category from previous sections
to train Llama-3.1:8b with a single epoch. Ground-truth labels serve as a reference (\textit{RLHF}). Both GRPO and DPO optimization algorithms are used. Results are shown in Tables \ref{tab:rl_annotators} and \ref{tab:rl_annotators_time}.

\begin{table}
\centering
\setlength{\tabcolsep}{4pt}
\begin{tabular}{|c|rc|c|rr|c|rr|}

\cline{4-9}
\multicolumn{3}{c|}{} & \multicolumn{3}{c|}{GRPO} & \multicolumn{3}{c|}{DPO}\\

\cline{2-9}
\multicolumn{1}{c|}{} & UMC & PT & GT & LLM & SLM & GT & LLM & SLM\\
\hline
\hline
D2 & 0.79 & \textbf{0.88} & 0.75 & 0.81 & 0.59 & 0.56 & 0.72 & 0.50 \\
D3 & 0.49 & 0.56 & 0.52 & 0.55 & 0.48 & 0.41 & \textbf{0.58} & 0.41 \\
D4 & \textbf{0.98} & 0.96 & 0.96 & 0.97 & 0.93 & 0.86 & 0.94 & 0.63 \\
D5 & 0.40 & 0.63 & 0.77 & 0.76 & 0.76 & \textbf{0.86} & 0.85 & 0.69 \\
D6 & 0.17 & 0.49 & 0.78 & 0.75 & 0.80 & 0.84 & \textbf{0.85} & 0.72 \\
D7 & 0.19 & 0.34 & 0.66 & 0.61 & 0.73 & \textbf{0.76} & 0.72 & 0.63 \\
D8 & 0.20 & 0.33 & 0.48 & 0.36 & 0.60 & 0.61 & \textbf{0.73} & 0.57 \\
D9 & \textbf{0.89} & 0.86 & 0.88 & \textbf{0.89} & 0.84 & 0.76 & 0.83 & 0.62 \\
\hline
Mean & 0.51 & 0.63 & 0.73 & 0.71 & 0.71 & 0.71 & \textbf{0.78} & 0.60 \\
\hline
\end{tabular}

\caption{F1-score for Reinforcement Learning on Llama-3.1:8b on testing data for different teachers compared with baselines. LLM is Qwen-2.5:32b and SLM is RoBERTa on Qwen-2.5:32b labels. {\color{black}PT stands for Pre-trained.}}
\label{tab:rl_annotators}
\end{table}






\begin{table}
\small
\begin{tabular}{|c|cc||rr|rr|}
\cline{4-7}
\multicolumn{3}{c||}{} & \multicolumn{4}{c|}{RL LLM} \\
\cline{4-7}
\multicolumn{3}{c||}{} & \multicolumn{2}{c|}{GRPO} & \multicolumn{2}{c|}{DPO} \\
\cline{2-7}
\multicolumn{1}{c|}{}& Stud. & Teac. & LLM & SLM & LLM & SLM \\
\hline
\hline
Ann. & 0 & 0 & 5,219 & 1,364 & 5,219 & 1,364  \\
Train & 0 & 0 & 18,071 & 17,562 & 5,453 & 5,644 \\
\hline
D2 & 1,495 & 2,801 & 375 & 347 & 1,209 & 3,129 \\
D3 & 2,151 & 3,329 & 504 & 470 & 2,450 & 4,392 \\
D4 & 2,118 & 4,663 & 724 & 669 & 1,124 & 3,344 \\
D5 & 7,099 & 7,135 & 1,168 & 1,074 & 3,571 & 8,443 \\
D6 & 8,446 & 6,628 & 1,069 & 993 & 4,396 & 9,746 \\
D7 & 7,562 & 8,264 & 1,571 & 1,450 & 7,729 & 10,928 \\
D8 & 10,306 & 5,898 & 925 & 856 & 4,145 & 7,925 \\
D9 & 3,111 & 5,473 & 792 & 726 & 2,127 & 5,044 \\
\hline
Sum & 42,291 & 44,195 & 30,423 & 25,515 & 37,427 & 59,962 \\
\hline
\end{tabular}

\caption{{\color{black} Times (seconds) for Reinforcement Learning on Llama-3.1:8b on testing data for different teachers compared with baselines. LLM is Qwen-2.5:32b and SLM is RoBERTa on Qwen-2.5:32b labels.}}
\label{tab:rl_annotators_time}
\end{table}

Regarding effectiveness, GRPO with noisy labels achieves slightly lower F1 (0.71) than with ground-truth labels (0.73), as expected due to noise. Surprisingly, training with DPO and LLM-generated labels achieves a higher F1 (0.78) than with human ground-truth (0.71), despite being noisier. {\color{black} This unexpected result may stem from better alignment of LLM-style preferences with the DPO objective, though further investigation is needed}. The combination of DPO with SLM annotator achieves only~0.60, confirming its limited capabilities.

Regarding time efficiency {\color{black} in Table \ref{tab:rl_annotators_time}}, GRPO requires more training time, because it generates multiple candidates per example. DPO trains faster but incurs higher inference costs: 
{\color{black} inference when trained from the LLM teacher is approximately twice as fast as when trained from the SLM teacher, since the LLM provides higher-quality preference signals, leading to better-aligned decision boundaries and reducing the number of uncertain cases processed during inference.}

{\color{black}
\textit{Summary: DPO with the LLM annotator is the most effective strategy and trains faster than GRPO, but incurs higher inference costs due to preference-based scoring.}
}






\textbf{Model Version}: We evaluate whether the underlying model version affects RL performance. Pre-trained and fine-tuned models (from Section~\ref{sec:eval_sft} with Qwen-2.5:32b noisy labels) are used with a single epoch and LLM labels. The results are in Table~\ref{tab:rl_model}.

\begin{table}
\centering

\begin{tabular}{|c|c|cc||cc|cc|}

\cline{3-8}
\multicolumn{2}{c|}{} & \multicolumn{2}{c||}{Llama} & \multicolumn{2}{c|}{GRPO} & \multicolumn{2}{c|}{DPO}\\

\cline{2-8}
\multicolumn{1}{c|}{} & UMC & PT & SFT & PT & SFT & PT & SFT\\
\hline
\hline
D2 & 0.79 & 0.88 & 0.88 & 0.81 & 0.90 & 0.72 & 0.89 \\
D3 & 0.49 & 0.56 & 0.65 & 0.55 & 0.65 & 0.58 & 0.58 \\
D4 & 0.98 & 0.96 & 0.99 & 0.97 & 0.98 & 0.94 & 0.93 \\
D5 & 0.40 & 0.63 & 0.91 & 0.76 & 0.90 & 0.85 & 0.86 \\
D6 & 0.17 & 0.49 & 0.90 & 0.75 & 0.87 & 0.85 & 0.86 \\
D7 & 0.19 & 0.34 & 0.77 & 0.61 & 0.76 & 0.72 & 0.72 \\
D8 & 0.20 & 0.33 & 0.82 & 0.36 & 0.75 & 0.73 & 0.71 \\
D9 & 0.89 & 0.86 & 0.88 & 0.89 & 0.92 & 0.83 & 0.87 \\
\hline
Mean & 0.51 & 0.63 & 0.85 & 0.71 & 0.84 & 0.78 & 0.80 \\
\hline
\end{tabular}

\caption{F1-score for Reinforcement Learning on Llama-3.1:8b on testing data for different versions of model compared with baselines. PT stands for Pre-trained and SFT for Supervised Fine-Tuning.}
\label{tab:rl_model}
\end{table}

Both GRPO and DPO improve over the pre-trained model (with an 18\% and a 2\% increase in F1, respectively). The larger gain for GRPO may result from reward functions encouraging concise and precise answers, whereas DPO relies on more informative examples and requires less effort. Nonetheless, all RL-based methods still underperform compared to the original fine-tuned model without RL {\color{black}(Llama-SFT)}, which remains superior on average. 

{\color{black} \textbf{Discussion}}: {\color{black} While {\color{black}all methods examined in this experimental analysis} are affected by the noisy labels generated by the Teacher, GRPO is additionally dependent on the quality of the reward functions. We can safely assume that the designed reward functions could be improved, leaving room for future work. Regarding DPO, we note that the counter-examples used in our scenario are all blocking options. Alternatively, counter-examples could be selected to be more distinct, e.g., a random entity from the collection or a blocking option with a ranking much lower than the original $k$ candidates. Since our scenario follows an end-to-end pipeline, we wanted the model to learn to separate the two classes solely based on blocking options; however, given the performance of the RL-based models, this appears to be particularly challenging, especially considering that the annotated answer may not be correct in the first place.} 

\textit{Summary: Reinforcement Learning improves pre-trained models, but provides limited benefit over already fine-tuned models.}

\subsection{Explanations}
\noindent \textbf{Q5: How can we create explanations instead of single answers using Fine-Tuning and Reinforcement Learning?}







\begin{table}
\centering
\setlength{\tabcolsep}{4.4pt}
\begin{tabular}{|c|cc||cc|ccc|}

\cline{4-8}
\multicolumn{3}{c|}{} & \multicolumn{2}{c|}{SFT} & \multicolumn{3}{c|}{DPO}\\
\cline{2-8}
\multicolumn{1}{c|}{} & UMC & PT & An. & Ex. & An. & Ex.-PT & Ex.-SFT \\
\hline
\hline
D2 & 0.79 & 0.88 & 0.88 & 0.83 & 0.72 & 0.11 & 0.08 \\
D3 & 0.49 & 0.56 & 0.65 & 0.63 & 0.58 & 0.27 & 0.28 \\
D4 & 0.98 & 0.96 & 0.99 & 0.98 & 0.94 & 0.14 & 0.04 \\
D5 & 0.40 & 0.63 & 0.91 & 0.87 & 0.85 & 0.52 & 0.64 \\
D6 & 0.17 & 0.49 & 0.90 & 0.89 & 0.85 & 0.61 & 0.82 \\
D7 & 0.19 & 0.34 & 0.77 & 0.77 & 0.72 & 0.64 & 0.82 \\
D8 & 0.20 & 0.33 & 0.82 & 0.83 & 0.73 & 0.64 & 0.71 \\
D9 & 0.89 & 0.86 & 0.88 & 0.88 & 0.83 & 0.13 & 0.07 \\
\hline
Mean & 0.51 & 0.63 & 0.85 & 0.84 & 0.78 & 0.38 & 0.43 \\
\hline
\end{tabular}

\caption{
{\color{black}
F1-score when integrating the explanations of the labels (Ex.) into SFT and RL with DPO in comparison to their answer-only counterparts (An.) and the two main baselines: UMC and Pre-trained (PT)}}
\label{tab:explanations}
\end{table}

In this experiment, we explore the use of explanations generated by the teacher model, instead of relying solely on single answers produced during labeling. We use the best teacher, Qwen-2.5:32b, to generate explanations for all candidate choices per query entity, including the \texttt{None} option, since the correct answer might not be among the candidates. These explanations are then used for both SFT and RL with DPO.
{\color{black}In this evaluation, we measure whether the model predicts correctly when using explanations, assessing whether the reasoning mode enhances performance. However, evaluating the content of the explanations remains an open yet promising task. From our indicative examination of the generated explanations, we observed that, when a Teacher provides a justification for an incorrect choice (Option X) while the correct choice is Option Y, it often simultaneously explains why Option Y could be correct. This demonstrates the model’s robustness against disorientation.}

{\color{black} Both SFT and DPO are strategies that rely on training with label examples: SFT provides the correct response, while DPO provides both the preferred response and rejectable alternatives. GRPO, however, is not applicable here, since efficiently evaluating free-text explanations is challenging.} The models are compared against their counterparts trained on single answers and against the baselines of UMC and pre-trained models. The results are reported in Table \ref{tab:explanations}.


We observe that SFT maintains similar performance when trained on either explanations or single answers, with the generated responses remaining coherent and meaningful. In contrast, DPO shows a clear performance drop: pre-trained models lose about 50\% of their effectiveness, while fine-tuning offers little improvement. {\color{black} This likely occurs because DPO overfits to the writing style or length of the explanations instead of focusing on their correctness, especially when explanations include reasoning for wrong answers.} Exceptions appear in the movie datasets (D6–D8), where the longer and more descriptive explanations seem to fit the DPO objective better, giving results similar to training with single answers.

\textit{Summary: {\color{black} Utilizing generated explanations to lead to correct answers is an effective, alternative solution {\color{black} when} using }Supervised Fine-Tuning, whereas DPO is not robust to this setting.}

\begin{table}
\centering
\setlength{\tabcolsep}{4pt}
\small
\begin{tabular}{|c|cc|c|cc|ccc|}
\cline{2-9}

\multicolumn{1}{c}{} & \multicolumn{2}{|c}{DistillER} & \multicolumn{6}{|c|}{SotA} \\

\cline{4-9}

\multicolumn{1}{c}{} & \multicolumn{2}{|c}{SFT on LLM} & \multicolumn{1}{|c}{LLM-Rel.} & \multicolumn{2}{|c}{SLM-Unsup.} &  \multicolumn{3}{|c|}{SLM-Sup.}  \\

\cline{2-9}
\multicolumn{1}{c|}{} & LLM & SLM & CEM & ZER & CBM & SWO & HGT & UCN \\
\hline
\hline
D2 & \textbf{0.88} & 0.48 & 0.47 & 0.45 & 0.00 & 0.72 & 0.59 & 0.52 \\
D3 & \textbf{0.65} & 0.41 & 0.41 & 0.26 & 0.33 & 0.50 & 0.29 & 0.50 \\
D4 & \textbf{0.99} & 0.85 & 0.83 & 0.94 & 0.86 & 0.28 & 0.98 & 0.60 \\
D5 & \textbf{0.91} & 0.89 & 0.50 & 0.22 & 0.43 & 0.36 & 0.58 & 0.68 \\
D6 & \textbf{0.90} & \textbf{0.90} & 0.19 & 0.81 & 0.48 & 0.08 & 0.41 & 0.77 \\
D7 & 0.77 & 0.79 & 0.25 & 0.55 & 0.40 & 0.20 & 0.52 & \textbf{0.96} \\
D8 & 0.82 & 0.79 & 0.08 & 0.57 & 0.09 & 0.39 & 0.53 & \textbf{0.86} \\
D9 & 0.88 & 0.65 & 0.68 & 0.86 & 0.68 & 0.24 & 0.84 & \textbf{0.92} \\
\hline
Mean & \textbf{0.85} & 0.72 & 0.43 & 0.58 & 0.41 & 0.35 & 0.59 & 0.73 \\
\hline
\end{tabular}

\caption{{\color{black} F1-score for DistillER against SotA algorithms. LLM is Qwen-2.5:32b and SLM is RoBERTa on Qwen-2.5:32b labels.}}
\label{tab:sota}
\end{table}

\subsection{Baselines}

\textbf{Q6: How does DistillER compare to existing unsupervised and supervised methods in Supervised Fine-Tuning?}

In this experiment, we compare DistillER against state-of-the-art ER methods across different categories. From the LLM-based approaches, we consider ComEM (\textbf{CEM})~\cite{DBLP:journals/corr/abs-2405-16884}, an unsupervised method that filters candidates with MATCH prompts before applying SELECT prompts on a pre-trained Llama-3.1:8b. The second category {\color{black}of baseline methods} involves SLM models. {\color{black}From the unsupervised methods, we consider}: ZeroER (\textbf{ZER})~\cite{wu2020zeroer}, which models candidate pairs with similarity-based feature vectors under a Gaussian Mixture Model and CollaborEM (\textbf{CBM})~\cite{ge2021collaborem}, which employs multi-feature collaboration through automatic label generation and collaborative training.
For supervised methods, we 
{\color{black}apply} the noisy labels from Qwen-2.5:32b to the following: HierGAT (\textbf{HGT})~\cite{yao2022entity}, which captures interdependence between ER decisions and attribute relationships with a Hierarchical Graph Attention Transformer, SudoWoodo (\textbf{SWO})~\cite{wang2023sudowoodo}, which uses contrastive representation learning for similarity-aware data representations, and Unicorn (\textbf{UCN})~\cite{DBLP:journals/pacmmod/TuFTWL0JG23} in its supervised variant, fine-tuned on the specific training dataset. We compare all methods based on effectiveness (F1-score), following common practice in the literature. The results can be seen in Table \ref{tab:sota}.

{\color{black}
Starting with the LLM baseline, CEM shows very poor performance with the average F1-score amounting to just 0.43. The reason is its aggressive candidate filtering, which leaves some queries without any candidates, excluding them from the results. 

Poor performance is also exhibited by the SLM-based methods in most cases; a core limitation is that blocked pairs are often highly similar, offering insufficient distinction for reliable separation. Regarding unsupervised methods, although CBM achieves high Recall across most datasets, it suffers from low Precision, leading to poor F1-scores. This imbalance is likely due to the skewed training data, which favors the negative class, biasing the model toward predicting negatives. This effect is exacerbated in D2, where no positive predictions are made. Similarly poor performance is exhibited by most supervised models. For example, SWO, which leverages Contrastive Learning, cannot differentiate the two classes. Only UCN adapts well under these conditions, benefiting from the multi-task training and the broad training data, especially in D7-D9, where it outperforms DistillER. 

The superiority of DistillER 
over the SLM-based methods is not surprising, as the 
latter rely on smaller models, whereas the former leverages LLMs with stronger representations. DistillER achieves the best performance when fine-tuned with LLM-generated labels, but even when trained on SLM-generated labels, it remains competitive with the other state-of-the-art methods. 


\textit{Summary: DistillER yields state-of-the-art performance when fine-tuned with LLM-generated labels, and remains highly competitive even with SLM-generated labels, whereas the smaller SLM-based methods naturally face limitations.}
}

\section{Conclusions}
\label{sec:conclusions}

In this work, we studied Knowledge Distillation in Entity Resolution using LLMs and suggested a framework spanning three dimensions: Data Selection, Knowledge Elicitation, and Distillation Algorithms. For Data Selection, we proposed and evaluated several methods, concluding that ranking tuples by their maximum blocking similarity provides the best balance between effectiveness and efficiency. For Knowledge Elicitation, we compared LLM, SLM, and multi-LLM teachers, showing that LLM teachers can produce high-quality labels even with medium-sized models (e.g., 32B), while SLM teachers serve as a fast and efficient alternative. In Distillation, we evaluated two strategies: SFT and RL.
For SFT, the best results are obtained by fine-tuning LLMs on noisy labels generated by LLM teachers, whereas SLMs are less competitive. For RL, we applied established optimization algorithms and found that DPO on a fine-tuned model performs best, although it still falls short of the performance achieved by direct SFT with noisy labels. We also explored generating explanations instead of single-label answers, showing that SFT on LLM-generated explanations is the most effective strategy. Finally, when compared to state-of-the-art ER methods, DistillER with SFT achieves dominant performance, surpassing both LLM- and SLM-based approaches, regardless of whether labeled data are required.

\section*{Acknowledgements}
This work was partially funded by the EU projects STELAR (101070122), WiseFood (101181895) and AI-DAPT (101135826).  

\section*{Artifacts}
All of our code, datasets and fine-tuned models are publicly available at \url{https://github.com/alexZeakis/DistillER}.

\bibliographystyle{ACM-Reference-Format}
\bibliography{references}

\end{document}